\documentclass{article}
\usepackage{amsmath}
\usepackage{amssymb}
\usepackage[section]{algorithm}
\usepackage{numinsec}

\input{tcilatex}
\begin{document}

\title{Application of simplest random walk algorithms for pricing barrier
options}
\author{M. Krivko\thanks{%
Department of Mathematics, University of Leicester, Leicester LE1~7RH, UK;
E-mail: {\small mk211@le.ac.uk}} \and M.V. Tretyakov\thanks{%
School of Mathematical Sciences, University of Nottingham, Nottingham NG7
2RD, UK; E-mail Michael.Tretyakov@nottingham.ac.uk}}
\maketitle

\begin{abstract}
We demonstrate effectiveness of the first-order algorithm from [Milstein,
Tretyakov. Theory Prob. Appl. 47 (2002), 53-68] in application to barrier
option pricing. The algorithm uses the weak Euler approximation far from
barriers and a special construction motivated by linear interpolation of the
price near barriers. It is easy to implement and is universal: it can be
applied to various structures of the contracts including derivatives on
multi-asset correlated underlyings and can deal with various type of
barriers. In contrast to the Brownian bridge techniques currently commonly
used for pricing barrier options, the algorithm tested here does not require
knowledge of trigger probabilities nor their estimates. We illustrate this
algorithm via pricing a barrier caplet, barrier trigger swap and barrier
swaption.


\end{abstract}

\section{Introduction}

Barrier option contracts are among the most traded and oldest exotic
derivatives. They accommodate investors' view about the future market
behavior more closely and they are generally cheaper than the corresponding
plain vanilla options. Typically, a barrier option is activated (knocked in)
or deactivated (knocked out) depending on whether a vector of underlying
assets or their functional has crossed a specified barrier level, which
itself can be a functional of the underlying assets. Due to its attractive
features, barrier optionality has been introduced in a wide range of
derivatives products. In the context of credit risk, the event of default of
some reference entity can be modelled as a lower barrier on the equity of
the entity. This is the key idea in structural models for pricing popular
credit instruments such as credit default swaps (CDSs) and credit default
obligations (CDOs). More recently, barrier optionality has been also used to
model contingent convertible (CoCo) bonds which were introduced to provide
financial institutions with sufficient capital in times of distress or
systemic risk.

Closed form solutions for barrier option prices can be obtained only in some
particular settings. For instance, they are available in the case of a
single underlying asset and a constant barrier within the standard
Black-Scholes setup (see, e.g. \cite{Ikeda,Reiner,BrigoMercurio,Rebo,Schoe}%
). In products involving a large number of dependent assets numerical
approximation for pricing and hedging barrier options is usually inevitable
and this can be a challenging problem.

In this paper we assume that underlying assets are modelled via
multidimensional stochastic differential equations (SDEs) and we consider
European-type barrier options. The arbitrage price $u(t,x)$ of such an
option solves the Dirichlet problem for a linear parabolic partial
differential equation. Finding this price numerically requires efficient
weak approximations of diffusions in a bounded domain.

\textquotedblleft Ordinary\textquotedblright\ numerical methods for SDEs on
a finite time interval in $\mathbf{R}^{d}$ are based on a time
discretization \cite{KLP92,GN,MT1}. They ensure smallness of time increments
at each step, but might not ensure smallness of space increments. In \cite%
{GNR93,GN95b,MT02} (see also \cite{MT1}) a number of weak-sense
approximations for SDEs in a bounded domain were proposed, in which space
increments are controlled at each step so that the constructed approximation
belongs to the bounded domain. Approximations of \cite{GN95b} (see also \cite%
{MT1}) are based on adaptive control of a time step of numerical integration
of the SDEs. A step is chosen such that (of course, aside of reaching a
required accuracy) the next state of a Markov chain approximating in the
weak sense the SDEs' solution remains in the bounded domain with probability
one. This leads to a decrease of the time step when the chain is close to
the boundary of the domain. The chain is stopped in a narrow zone near the
boundary so that values of the solution $u(t,x)$ (i.e., the option price at
time $t$ and underlyings' price $x)$ in this zone can be approximated
accurately by the known values of the function $\varphi $ on the boundary
(i.e., the value of the option at the barrier). Another type of
approximations was proposed in \cite{MT02} (see also \cite{MT1}). In the
algorithm of weak order one from \cite{MT02} the step of numerical
integration of the SDEs is constant for points belonging to a certain time
layer $t=t_{k}.$ Far from the boundary, a Markov chain approximating the
SDEs' solution is constructed using the weak Euler scheme (i.e., using
discrete random variables for approximating the Wiener increments). When a
point is close to the boundary, we make an intermediate (auxiliary) step of
the random walk, which preserves the point in the time layer $t=t_{k}.$ On
this auxiliary step we \textquotedblleft flip a coin\textquotedblright\ to
decide whether to terminate the chain on the boundary or jump back in the
domain and continue the random walk. The construction of this step is based
on the idea of linear interpolation for the solution $u(t,x).$ The algorithm
is efficient and very easy to implement. Its simpler version of order $1/2$
in the weak sense is also presented in \cite{MT02,MT1}. In Section~\ref%
{secalg} we recall these two algorithms from \cite{MT02,MT1}. Due to our
knowledge, despite simplicity of these weak schemes, they have not been used
in financial applications. In this paper we try to fill this gap and
illustrate their applicability to pricing barrier options.

Currently, the popular numerical approach for pricing barrier options
exploits the Brownian bridge technique \cite%
{Andersen,Baldi,Gobet2000,Shevchenko,Zhou} (also see \cite{Gobet2009} for a
review and the references therein). It is based on simulation of a
one-dimensional Brownian bridge extremum between time steps and computing
analytically the associated probability of exiting the spatial domain for
each time interval of the partition. It was proved in \cite{Gobet2000} that
this approach realized along with the Euler scheme (which uses Gaussian
random variables for simulating Wiener increments) results in an
approximation of weak order one. The Brownian bridge technique relies on
analytical formulas for trigger probabilities and can run into difficulties
in the case of multiple barriers and/or correlated structure of the
underlyings when there are no closed formulas for the distribution of
extremum. Though some extensions to these more general and not uncommon
problems have been considered, e.g. in \cite{Gobet2009,Shevchenko}.

In contrast the simplest random walk algorithm of \cite{MT02} displays a
high degree of flexibility and can be applied to various structures of the
contracts including derivatives on multi-asset correlated underlyings and
can deal with various type of barriers, e.g. single, double and time
dependent barriers. In comparison with the Brownian bridge techniques the
method of \cite{MT02} does not require the knowledge of the trigger
probabilities nor their estimates.

In Sections~\ref{sec_cap}-\ref{sec_swpt} we present three examples on how to
apply the algorithm from \cite{MT02} for valuation of barrier options. These
contracts cover the most common types of barrier options. In the first
example (Section~\ref{sec_cap}), we consider an algorithm for barrier
derivatives where the payoff depends on a single underlying. As an
illustration, we deal with pricing a barrier caplet and provide
ready-for-implementation procedure which can be easily applied to similar
other contracts. The second example (Section~\ref{sec_trig}) is devoted to
multi-asset options with barriers imposed on all or some of the correlated
underlying assets. We illustrate this case by pricing a trigger swap. The
last example (Section~\ref{sec_swpt}) is barrier contracts written on an
asset that can be expressed through some other multi-asset underlying. As a
specific case, we consider valuation of a barrier swaption under the LIBOR
market model (LMM).

\section{Simplest random walks for stopped diffusions\label{secalg}}

Let $(\Omega ,\mathcal{F},\mathbb{P})$ be a complete probability space, $%
\mathcal{F}_{t},\ 0\leq t\leq T,$ be a filtration satisfying the usual
hypotheses, $(w_{t},\mathcal{F}_{t})$ be an $r$-dimensional standard Wiener
process. Let $G$ be a bounded domain in $\mathbf{R}^{d}$ and $%
Q=[t_{0},T)\times G$ be a cylinder in $\mathbf{R}^{d+1},$ $\Gamma =\bar{Q}%
\backslash Q$ be the part of the cylinder's boundary consisting of the upper
base and lateral surface. Price of barrier options with underlying modelled
by a diffusion process can usually be expressed as 
\begin{equation}
u(t,x)=E\left[ \varphi (\tau ,X_{t,x}(\tau ))Y_{t,x,1}(\tau
)+Z_{t,x,1,0}(\tau )\right] \,,  \label{Ha03}
\end{equation}%
where $X_{t,x}(s),$ $Y_{t,x,y}(s),$ $Z_{t,x,y,z}(s),$ $s\geq t,$ is the
solution of the Cauchy problem for the system of SDEs: 
\begin{eqnarray}
dX &=&(b(s,X)-\sigma (s,X)\mu (s,X))\,ds+\sigma (s,X)\,dw(s),\;\;X(t)=x,
\label{Ha04} \\
dY &=&c(s,X)Y\,ds+\mu ^{\intercal }(s,X)Y\,dw(s),\;\;Y(t)=y,  \label{Ha05} \\
dZ &=&g(s,X)Y\,ds+F^{\intercal }(s,X)Y\,dw(s),\;\;Z(t)=z,  \label{Ha06}
\end{eqnarray}%
$(t,x)\in Q,$ and $\tau =\tau _{t,x}$ is the first exit time of the
trajectory $(s,X_{t,x}(s))$ to the boundary $\Gamma .$ In (\ref{Ha04})-(\ref%
{Ha06}), $b(s,x)$ is a $d$-dimensional column-vector$,$ the $\sigma (s,x)$
is a $d\times r$ matrix, $\mu (s,x)$ and $F(s,x)$ are $r$-dimensional
vectors, and $Y(s),$ $Z(s),$ $c(s,X)$ and $g(s,X)$ are scalars. We assume
that all the coefficients in (\ref{Ha04})-(\ref{Ha06}), the function $%
\varphi (t,x)$ defined on $\Gamma $ and the boundary $\partial G$ of the
space domain $G$ satisfy some regularity conditions.

We note that the value of the expectation $u(t,x)$ in (\ref{Ha03}) does not
depend on a choice of functions $\mu (s,x)$ and $F(s,x).$ This flexibility
can be used for reducing variance of the random variable $\varphi (\tau
,X_{t,x}(\tau ))Y_{t,x,1}(\tau )+Z_{t,x,1,0}(\tau )$ with the aim of
reducing the statistical error in computing $u(t,x)$ via the Monte Carlo
technique \cite{GN,MT1}. For instance, if $\mu $ and $F$ are such that 
\begin{equation}
\sum_{i=1}^{d}\sigma ^{ij}\frac{\partial u}{\partial x^{i}}+u\mu
^{j}+F^{j}=0,\;j=1,\ldots ,r,  \label{De29}
\end{equation}%
then $Var[\varphi (\tau ,X_{t,x}(\tau ))Y_{t,x,1}(\tau )+Z_{t,x,1,0}(\tau
)]=0$ and $\varphi (\tau ,X_{t,x}(\tau ))Y_{t,x,1}(\tau )+Z_{t,x,1,0}(\tau
)\equiv u(t,x)$ \cite{MS,MT1}. As we see from (\ref{De29}), optimal $\mu $
and $F$ require knowledge of the the solution $u(t,x)$ (i.e., the option
price) to the considered problem and its derivatives (i.e., deltas) which is
impractical. However, instead of the exact $u(t,x)$ in (\ref{De29}), one can
use its approximation (e.g. price and deltas for a related option for which
the closed-form solution is known) to find some suboptimal $\mu $ and $F$
which can lead to variance reduction \cite{Glasserman,vari}.

To simulate (\ref{Ha03})-(\ref{Ha06}), we need an approximation of the
trajectory $(s,X(s))$ which satisfies some restrictions related to its
nonexit from the domain $\bar{Q}.$ Let us recall two algorithms for (\ref%
{Ha03})-(\ref{Ha06}) from \cite{MT02,MT1}.

We apply the weak explicit Euler approximation with the simplest simulation
of noise to the system (\ref{Ha04})-(\ref{Ha06}): 
\begin{eqnarray}
X_{t,x}(t+h) &\approx &X=x+h\left( b(t,x)-\sigma (t,x)\,\mu (t,x)\right)
+h^{1/2}\sigma (t,x)\,\xi \,,\;\;\;\;  \label{Hc01} \\
Y_{t,x,y}(t+h) &\approx &Y=y+hc(t,x)\,y+h^{1/2}\mu ^{\intercal
}(t,x)\,y\,\xi \,,  \label{Hc02} \\
Z_{t,x,y,z}(t+h) &\approx &Z=z+hg(t,x)\,y+h^{1/2}F^{\intercal }(t,x)\,y\,\xi
\,,  \label{Hc03}
\end{eqnarray}%
where $h>0$ is a time-discretization step (a sufficiently small number), $%
\xi =(\xi ^{1},\ldots ,\xi ^{r})^{\intercal },$ $\xi ^{i},$ $i=1,\ldots ,r,$
are mutually independent random variables taking the values $\pm 1$ with
probability $1/2.$ Clearly, the random vector $X$ takes $2^{r}$ different
values.

Introduce the set of points close to the boundary (a boundary zone) $%
S_{t,h}\subset \bar{G}$ on the layer $t:$ we say that $x\in S_{t,h}$ if at
least one of the $2^{r}$ values of the vector $X$ is outside $\bar{G}.$ It
is not difficult to see that due to compactness of $\bar{Q}$ there is a
constant $\lambda >0$ such that if the distance from $x\in G$ to the
boundary $\partial G$ is equal to or greater than $\lambda \sqrt{h}$ then $x$
is outside the boundary zone and, therefore, for such $x$ all the
realizations of the random variable $X$ belong to $\bar{G}.$

Since we should impose restrictions on an approximation of the system (\ref%
{Ha04}) so that it does exit from the domain $\bar{G}$, the formulas (\ref%
{Hc01})-(\ref{Hc03}) can be used only for the points $x\in \bar{G}\backslash
S_{t,h}$ on the layer $t,$ and a special construction is required for points
from the boundary zone. Let $x\in S_{t,h}.$ Denote by $x^{\pi }\in \partial
G $ the projection of the point $x$ on the boundary of the domain $G$ (the
projection is unique assuming that $h$ is sufficiently small and $\partial G$
is smooth) and by $n(x^{\pi })$ the unit vector of internal normal to $%
\partial G$ at $x^{\pi }.$ Introduce the random vector $X_{x,h}^{\pi }$
taking two values $x^{\pi }$ and $x+h^{1/2}\lambda n(x^{\pi })$ with
probabilities $p=p_{x,h}$ and $q=q_{x,h}=1-p_{x,h},$ respectively, where 
\begin{equation}
p_{x,h}=\frac{h^{1/2}\lambda }{|x+h^{1/2}\lambda n(x^{\pi })-x^{\pi }|}\,.
\label{prob}
\end{equation}%
This construction is motivated by the following observation \cite{MT02}. If $%
v(x)$ is a twice continuously differentiable function with the domain of
definition $\bar{G},$ then an approximation of $v(x)$ by the expectation $%
Ev(X_{x,h}^{\pi })$ corresponds to linear interpolation and 
\begin{equation}
v(x)=Ev(X_{x,h}^{\pi })+O(h)=pv(x^{\pi })+qv(x+h^{1/2}\lambda n(x^{\pi
}))+O(h)\,.  \label{Hc04}
\end{equation}

We emphasize that the second value $x+h^{1/2}\lambda n(x^{\pi })$ does not
belong to the boundary zone. We also note that $p$ is always greater than $%
1/2$ (since the distance from $x$ to $\partial G$ is less than $%
h^{1/2}\lambda )$ and that if $x\in \partial G$ then $p=1$ (since in this
case $x^{\pi }=x).$

Let a point $(t_{0},x_{0})\in Q.$ We would like to find the value $%
u(t_{0},x_{0}).$\ Introduce a discretization of the interval $\left[ t_{0},T%
\right] ,$ for definiteness the equidistant one: 
\begin{equation*}
t_{0}<t_{1}<\cdots <t_{M}=T,\;\;h:=(T-t_{0})/M.
\end{equation*}

To approximate the solution of the system (\ref{Ha04}), we construct a
Markov chain $(t_{k},X_{k})$ which stops when it reaches the boundary $%
\Gamma $ at a random step $\varkappa \leq M.$ The resulting algorithm can be
formulated as Algorithm~\ref{Hca01} given below.

\begin{algorithm}[h,t,b]
\caption{Algorithm of weak order one for (\ref{Ha03})-(\ref{Ha06})}
\label{Hca01}$\;\;$

$ 
\begin{tabular}{ll}
\ \ STEP 0. & $\;\;X_{0}^{\prime }=x_{0},\;Y_{0}=1,\;Z_{0}=0,\;k=0.$\medskip \\ 
\begin{tabular}{l}
STEP 1. \\ 
$\ \ $\medskip \\ 
$\ \ $%
\end{tabular}
& 
\begin{tabular}{l}
If $X_{k}^{\prime }\notin S_{t_{k},h},$\ then $X_{k}=X_{k}^{\prime }$ and go
to STEP 3. \\ 
If $X_{k}^{\prime }\in S_{t_{k},h},$ \ then either $X_{k}=X_{k}^{\prime \pi
} $ \ with probability \\ 
$p_{X_{k}^{\prime },h}\;$or $X_{k}=X_{k}^{\prime }+h^{1/2}\lambda
n(X_{k}^{\prime \pi })\ \ \text{with}$\ probability$\;q_{X_{k}^{\prime
},h}\,.$\medskip%
\end{tabular}
\\ 
\begin{tabular}{l}
STEP 2. \\ 
$\;\;$\medskip%
\end{tabular}
& 
\begin{tabular}{l}
If\ \ $X_{k}=X_{k}^{\prime \pi }$, \ then STOP\ and\ $\varkappa =k,$ \\ 
$X_{\varkappa }=X_{k}^{\prime \pi },\;Y_{\varkappa }=Y_{k},\;Z_{\varkappa
}=Z_{k}\,.$\medskip%
\end{tabular}
\\ 
\begin{tabular}{l}
STEP 3. \\ 
$\ \ $\medskip \\ 
$\;\;$%
\end{tabular}
& 
\begin{tabular}{l}
Simulate $\xi _{k}$ \ and find $X_{k+1}^{\prime },\;Y_{k+1},\;Z_{k+1}$ \
according to  \\
$(\ref{Hc01})$-$(\ref{Hc03})$ for $t=t_{k},\;x=X_{k},\;y=Y_{k},\;z=Z_{k},$
\\ 
$\xi =\xi _{k}\,.$\medskip%
\end{tabular}
\\ 
\begin{tabular}{l}
STEP 4. \\ 
$\ \ $%
\end{tabular}
& 
\begin{tabular}{l}
If $k+1=M,$\ STOP and $\varkappa =M,$ $X_{\varkappa }=X_{M}^{\prime },$ $%
Y_{\varkappa }=Y_{M},$ \\ 
$Z_{\varkappa }=Z_{M},$ otherwise $k=k+1$ \ and return to STEP 1.%
\end{tabular}%
\end{tabular}
$
\end{algorithm}

It is proved in \cite{MT02} (see also \cite{MT1}) that under appropriate
regularity assumptions on the coefficients of (\ref{Ha04})-(\ref{Ha06}), the
boundary condition $\varphi (t,x)$ in (\ref{Ha03}) and on the boundary $%
\partial G$ Algorithm~\ref{Hca01} converges with weak order one.

The next algorithm is obtained by a simplification of Algorithm~\ref{Hca01}:
as soon as $X_{k}$ gets into the boundary domain $S_{t_{k},h},$ the random
walk terminates, i.e., $\varkappa =k,$ and $\bar{X}_{\varkappa }=X_{k}^{\pi
} $, $Y_{\varkappa }=Y_{k}$, $Z_{\varkappa }=Z_{k}$ is taken as the final
state of the Markov chain. The resulting algorithm takes the form of
Algorithm~\ref{Hca02}.

\begin{algorithm}[h,t,b]
\caption{Algorithm of weak order $1/2$ for (\ref{Ha03})-(\ref{Ha06})}
\label{Hca02}$\;\;$

$ 
\begin{tabular}{ll}
\ \ STEP 0. & $\;\;X_{0}=x_{0},\;Y_{0}=1,\;Z_{0}=0,\;k=0.$\medskip \\ 
\begin{tabular}{l}
STEP 1. \\ 
$\ \ $\medskip \\ 
$\ \ $%
\end{tabular}
& 
\begin{tabular}{l}
If $X_{k}\notin S_{t_{k},h},$\ then go to STEP 2. \\ 
If $X_{k}\in S_{t_{k},h},$ \ then STOP and\ $\varkappa =k,$ $\bar{X}%
_{\varkappa }=X_{k}^{\pi }$, \\ 
$Y_{\varkappa }=Y_{k}$, $Z_{\varkappa }=Z_{k}\,.$\medskip%
\end{tabular}
\\ 
\begin{tabular}{l}
STEP 2. \\ 
$\ \ $\medskip \\ 
$\;\;$%
\end{tabular}
& 
\begin{tabular}{l}
Simulate $\xi _{k}$ \ and find $X_{k+1},\;Y_{k+1},\;Z_{k+1}$ \ according to
\\ 
$(\ref{Hc01})$-$(\ref{Hc03})$ for $t=t_{k},\;x=X_{k},\;y=Y_{k},$ \\ 
$z=Z_{k},\;\xi =\xi _{k}\,.$\medskip%
\end{tabular}
\\ 
\begin{tabular}{l}
STEP 3. \\ 
$\ \ $%
\end{tabular}
& 
\begin{tabular}{l}
If $k+1=M,$\ STOP and $\varkappa =M,$ $\bar{X}_{\varkappa }=X_{M},$ $%
Y_{\varkappa }=Y_{M},$ \\ 
$Z_{\varkappa }=Z_{M},$ otherwise $k=k+1$ \ and return to STEP 1.%
\end{tabular}%
\end{tabular}
$
\end{algorithm}

It is proved in \cite{MT02,MT1} that under appropriate regularity
assumptions on the coefficients of (\ref{Ha04})-(\ref{Ha06}), the boundary
condition $\varphi (t,x)$ in (\ref{Ha03}) and on the boundary $\partial G$
Algorithm~\ref{Hca02} converges with weak order $1/2$. We note that in
one-dimension (i.e., in the case of a single underlying) Algorithm~\ref%
{Hca02} is analogous to pricing barrier options by binary trees (see, e.g. 
\cite{Der95}).

\section{LIBOR Market Model\label{secopt}}

We will now assume that there exists an arbitrage-free market with
continuous and frictionless trading taking place inside a finite time
horizon $\left[ t_{0},t^{\ast }\right] $.

Among the most important benchmark interest rates is the London Interbank
Offered Rate (LIBOR). It is based on simple (or simply compounded) interest.
The forward LIBOR rate $L(t,T,T+\delta )$ is the rate set at time $t$ for
the interval $[T,T+\delta ],$ $t\leq T.$ If we enter into a contract at time 
$t$ to borrow one unit at time $T$ and repay it with interest at time $%
T+\delta ,$ the interest due will be $\delta L(t,T,T+\delta ).$

A simple replication argument (see, e.g., \cite{BrigoMercurio}) relates
LIBOR rates and bond prices via the following identity 
\begin{equation}
L(t,T,T+\delta )=\frac{1}{\delta }\left( \frac{P(t,T)}{P(t,T+\delta )}%
-1\right) ,  \label{libor_bond}
\end{equation}%
where $P(t,T)$ is the price at time $t\leq T$ of a default-free zero coupon
bond.

For simplicity, we fix an equidistant finite set of maturities or tenor dates%
\begin{equation}
T_{0}<\cdots <T_{N}=T^{\ast },\ \ T_{i}=i\delta ,\ \ i=0,\ldots ,N,
\label{Tdis}
\end{equation}%
where%
\begin{equation*}
\delta =(T^{\ast }-T_{0})/N,
\end{equation*}%
denotes the fixed length of the interval between tenor dates.

Let us introduce a simplified notation for the time $t$ forward LIBOR\ rate
for the accrual period $\left[ T_{i},T_{i+1}\right] $ and the payment at $%
T_{i+1}$: 
\begin{eqnarray*}
L^{i}(t) &:&=L(t,T_{i},T_{i+1}), \\
t_{0} &\leq &t\leq t^{\ast }\wedge T_{i},\ \ t_{0}<T_{i}\leq T^{\ast },\text{
\ }i=0,\ldots ,N-1.
\end{eqnarray*}

In the case of LIBOR Market Model (LMM) the arbitrage-free dynamics of $%
L^{i}(t)$ under the forward measure $\mathrm{Q}^{T_{k+1}}$ associated with
the numeraire $P(t,T_{k+1})$ can be written as the following system of SDEs
(see, e.g. \cite{BrigoMercurio,Rebo,Schoe}): 
\begin{equation}
\frac{dL^{i}(t)}{L^{i}(t)}=\left\{ 
\begin{array}{c}
\sigma _{i}(t)\sum\limits_{j=k+1}^{i}\frac{\delta L^{j}(t)}{1+\delta L^{j}(t)%
}\rho _{i,j}\sigma _{j}(t)dt+\sigma _{i}(t)dW_{i}^{T_{k+1}}(t),\text{ }i>k,%
\text{ }t\leq T_{k}, \\ 
\sigma _{i}(t)dW_{i}^{T_{k+1}}(t),\text{ }i=k,\text{ }t\leq T_{i}, \\ 
-\sigma _{i}(t)\sum\limits_{j=i+1}^{k}\frac{\delta L^{j}(t)}{1+\delta
L^{j}(t)}\rho _{i,j}\sigma _{j}(t)dt+\sigma _{i}(t)dW_{i}^{T_{k+1}}(t),\text{
}i<k,\text{ }t\leq T_{i},%
\end{array}%
\right.  \label{sde}
\end{equation}%
where $W^{T_{k+1}}=(W_{0}^{T_{k+1}},\ldots ,W_{N-1}^{T_{k+1}})^{\top }$ is
an $N$-dimensional correlated Wiener process defined on a filtered
probability space $\left( \Omega ,\mathcal{F},\left\{ \mathcal{F}%
_{t}\right\} _{t_{0}\leq t\leq t^{\ast }},\mathrm{Q}^{T_{k+1}}\right) $; the
instantaneous correlation structure is defined as%
\begin{equation}
E\left[ W_{i}^{T_{k+1}}(t)W_{j}^{T_{k+1}}(t)\right] =\rho _{i,j},\text{ \ }%
i,j=0,\ldots ,N-1;  \label{corr}
\end{equation}%
and $\sigma _{i}(t),$ $i=0,\ldots ,N-1,$ are instantaneous volatilities
which we assume here to be deterministic bounded functions.

Let $\rho $ be the instantaneous correlation matrix with elements $\rho
_{i,j}.$ To simulate the correlated Wiener processes, we will use the
pseudo-root of the correlation matrix $\rho $ defined via the equation 
\begin{equation}
\rho =UU^{\intercal },  \label{U}
\end{equation}%
where $U$ is an upper triangular matrix. Using $U$ and introducing $N$%
-dimensional standard Wiener process, one can re-write (\ref{sde}) in the
form of (\ref{Ha04}).

In what follows we will assume that the current time $t_{0}$ is set to $0$.
For convenience, we also assume a unit notional value of all the contracts
we introduce below. In our numerical experiments in the next sections we
take the correlation function of the form:%
\begin{equation}
\rho _{i,j}=\exp (-\beta \left\vert T_{i}-T_{j}\right\vert ).  \label{rho}
\end{equation}

\section{Barrier cap/floor\label{sec_cap}}

In this section we consider Monte Carlo evaluation of barrier options
written on a single underlying. We use a knock-out caplet for illustration,
though our treatment is rather general and can be used to value different
barrier option, for instance European and Parisian barrier options and
options with different barriers including single, double and time-dependent
barriers, both for fixed-income and equity markets.

An Interest Rate Cap is a security that allows its holder to benefit from
low floating rates and be protected from high ones. Similarly, an Interest
Rate Floor is an instrument designed to protect from low floating interest
rates yet allow the holder to benefit from high rates. Formally, a cap price
is obtained by summing up the prices of the underlying caplets, call options
on successive LIBOR rates. Also, a floor is a strip of floorlets, put
options on successive LIBOR rates.

A knock-out caplet pays the same payoff as a regular caplet as long as a
prescribed barrier rate $H$ is not reached from below by the corresponding
LIBOR rate before the option expires. More specifically, the price at time $%
t\leq T_{0}$ of knock-out caplet set at time $T_{i-1}$ with payment date at $%
T_{i},$ $i\geq 1,$ with strike $K$ and unit cap nominal value is given by%
\begin{equation}
V_{caplet}(t)=\delta P(t,T_{i+1})E^{\mathrm{Q}^{T_{i+1}}}\left[ \left.
\left( L^{i}(T_{i})-K\right) _{+}\chi \left( \theta >T_{i}\right)
\right\vert \mathcal{F}_{t}\right] ,  \label{caplet price}
\end{equation}%
where $\theta $ is the first exit time of $L^{i}(s),$ $s\geq t,$ from the
interval $G=(0,H).$ Let $\tau $ be the first exit time of the space-time
diffusion $(s,L^{i}(s))$ from the domain $Q=[t,T_{i})\times (0,H).$
Obviously, $\tau =\theta \wedge T_{i}$.

The dynamics of $L^{i}(s)$ under $\mathrm{Q}^{T_{i+1}}$ is (see (\ref{sde})):%
\begin{equation}
\frac{dL^{i}(s)}{L^{i}(s)}=\sigma _{i}(s)dW_{i}^{T_{i+1}}(s),\text{ }s\leq
T_{i}.  \label{sde caplet}
\end{equation}%
Note that the correlation structure of (\ref{sde}) does not influence the
price of the knock-out caplet since it does not depend on the joint dynamics
of forward rates.

One can observe that the dynamics (\ref{sde caplet}) coincides with the
model of a stock price process under the risk-neutral measure in the case of
zero interest rate. This means that by dropping the factor $\delta
P(t,T_{i+1})$ in (\ref{caplet price}), the valuation of European barrier
options on equity with zero interest rate and of barrier caplets under the
LMM coincide.

In the considered case the price of the barrier caplet has the well-known
closed-form solution:%
\begin{eqnarray}
V_{caplet}(t) &=&V_{caplet}(t,L^{i}(t))  \label{caplet_exact} \\
&=&\delta P(t,T_{i+1})\left\{ L^{i}(t)\left[ \Phi (\delta
_{+}(L^{i}(t)/K,v_{i}))-\Phi (\delta _{+}(L^{i}(t)/H,v_{i}))\right] \right. 
\notag \\
&&-K\left[ \Phi (\delta _{-}(L^{i}(t)/K,v_{i}))-\Phi (\delta
_{-}(L^{i}(t)/H,v_{i}))\right]  \notag \\
&&-H\left[ \Phi (\delta _{+}(H^{2}/(KL^{i}(t)),v_{i}))-\Phi (\delta
_{+}(H/L^{i}(t),v_{i}))\right]  \notag \\
&&\left. +KL^{i}(t)\left[ \Phi (\delta _{-}(H^{2}/(KL^{i}(t)),v_{i}))-\Phi
(\delta _{-}(H/L^{i}(t),v_{i}))\right] /H\right\} ,  \notag
\end{eqnarray}%
where $\Phi (\cdot )$ denotes the standard normal cumulative distribution
function and%
\begin{equation}
\delta _{+}(x,v)=(\ln x+v^{2}/2)/v,\ \ \ \delta _{-}(x,v)=(\ln x-v^{2}/2)/v,
\label{d1}
\end{equation}%
and%
\begin{equation*}
v_{i}^{2}=\int_{t}^{T_{i}}\left( \sigma _{i}(s)\right) ^{2}ds.
\end{equation*}%
This analytical result will be used in our numerical experiments to access
the performance of proposed algorithms. We note that the algorithm presented
in this example can be easily extended to a more general model of underlying
when the closed-form solution might be not available. In particular, there
is no difficulty in including a drift term in the underlying dynamics (see
also Sections~\ref{sec_trig} and~\ref{sec_swpt}). In the experiments we
simulate 
\begin{equation}
\tilde{V}_{caplet}(t)=V_{caplet}(t)/\delta P(t,T_{i+1}),  \label{tildaV}
\end{equation}%
i.e. we drop $\delta P(t,T_{i+1})$ from (\ref{caplet price}), which does not
imply any loss of generality since the caplet price can easily be recovered
by multiplying $\tilde{V}_{caplet}(t)$ by the factor $\delta P(t,T_{i+1})$
observable at time $t$.

\subsection{Algorithm}

To preserve positivity of the LIBOR rate, we simulate the log dynamics
corresponding to (\ref{sde caplet}) rather than the LIBOR rate $L^{i}(t)$
itself. To illustrate the variance reduction technique discussed in Section~%
\ref{secalg}, we complement (\ref{sde caplet}) with the equation (cf. (\ref%
{Ha06})): 
\begin{equation}
dZ=F(s,L^{i})dW_{i}^{T_{i+1}}(s),\;\;Z(0)=0,  \label{zcap}
\end{equation}%
with (see (\ref{De29}) and (\ref{tildaV})) 
\begin{equation}
F(s,L^{i})=-\sigma _{i}(s)\frac{\partial }{\partial L^{i}}\tilde{V}%
_{caplet}(s,L^{i}(s))\ .  \label{fcap}
\end{equation}%
We choose a time step $h>0$ so that $M=T_{i}/h$ is an integer. We set $\ln
L_{0}^{i}=L^{i}(0)$ and $Z_{0}=0.$ The weak Euler scheme (\ref{Hc01}), (\ref%
{Hc03}) applied to (\ref{sde caplet}) in the log form and (\ref{zcap}) takes
the form: 
\begin{eqnarray}
\ln L_{k+1}^{i} &=&\ln L_{k}^{i}-\frac{1}{2}\left( \sigma _{i}(t_{k})\right)
^{2}h+\sigma _{i}(t_{k})\sqrt{h}\xi _{k+1},\text{ }  \label{euler caplet} \\
Z_{k+1} &=&Z_{k}+F(s,L_{k}^{i})\sqrt{h}\xi _{k+1}\ ,  \label{eulerz}
\end{eqnarray}%
where $\xi _{k}$ are independent random variables distributed by the law $%
P(\xi =\pm 1)=1/2.$

The boundary zone $S_{t,h}$ required for Algorithms~\ref{Hca01} and~\ref%
{Hca02}\ is chosen here as 
\begin{equation}
S_{t_{k},h}=\{L_{k}^{i}:\ \ln L_{k}^{i}\geq \ln H+\frac{1}{2}\sigma
_{i}^{2}(t_{k})h-\sigma _{i}(t_{k})\sqrt{h}\},  \label{capS}
\end{equation}%
i.e., the condition for $\ln L_{k+1}^{i}$ to be inside the domain $G$ is%
\begin{equation}
\ln L_{k}^{i}<\ln H+\frac{1}{2}\sigma _{i}^{2}(t_{k})h-\sigma _{i}(t_{k})%
\sqrt{h}\ ,  \label{cc2}
\end{equation}%
and the corresponding $\lambda _{k}$ for Algorithm~\ref{Hca01} is so that 
\begin{equation}
\lambda _{k}\sqrt{h}=-\frac{1}{2}\sigma _{i}^{2}(t_{k})h+\sigma _{i}(t_{k})%
\sqrt{h}\ .  \label{caplam}
\end{equation}%
We note that instead of (\ref{capS}) and (\ref{caplam}) we could take the
wider boundary zone $S_{t_{k},h}=\{L_{k}^{i}:\ \ln L_{k}^{i}\geq \ln
H-\sigma _{i}(t_{k})\sqrt{h}\}$ and correspondingly $\lambda _{k}=\sigma
_{i}(t_{k}).$ A wider boundary zone usually leads to a bigger numerical
integration error. In this example we cannot take a boundary zone narrower
than $S_{t_{k},h}$ in (\ref{capS}) because it would not ensure that the
chain $\ln L_{k}^{i}$ belongs to $\bar{G}.$

To realize Algorithm~\ref{Hca01}, we follow the random walk generated by (%
\ref{euler caplet}) and at each time $t_{k},$ we check whether at the next
step $L_{k+1}^{i}$ cannot cross the barrier $H$, i.e., we check whether the
condition (\ref{cc2}) holds. If it does, we perform (\ref{euler caplet})-(%
\ref{eulerz}) to find $\ln L_{k+1}^{i},$ $Z_{k+1}.$ Otherwise, $L_{k}^{i}$
has reached the boundary zone $S_{t_{k},h},$ where we make the auxiliary
step: we either stop the chain at $\ln H$ with probability $p:$ 
\begin{equation*}
p=\frac{\lambda _{k}\sqrt{h}}{\ln H-\ln L_{k}^{i}+\lambda _{k}\sqrt{h}}
\end{equation*}%
or we kick the current position of the random walk $\ln L_{k}^{i}$ back into
the domain to the position $\ln L_{k}^{i}-\lambda _{k}\sqrt{h}$ with
probability $1-p$ and then carry out (\ref{euler caplet})-(\ref{eulerz}) to
find $\ln L_{k+1}^{i}$, $Z_{k+1}.$ If $k+1=M,$ we stop, otherwise we
continue with the algorithm. The outcome of simulating each trajectory is a
point $(t_{\varkappa },\ln L_{\varkappa }^{i},Z_{\varkappa }).$

To realize Algorithm~\ref{Hca02}, we also follow the random walk generated
by (\ref{euler caplet}), and at each time $t_{k},$ we check whether the
condition (\ref{cc2}) holds. If it does not, $L_{k}^{i}$ has reached the
boundary zone $S_{t_{k},h}$ and we stop the chain at $\ln H$. If it does, we
perform (\ref{euler caplet})-(\ref{eulerz}) to find $\ln L_{k+1}^{i},$ $%
Z_{k+1}.$ If $k+1=M,$ we stop, otherwise we continue with the algorithm. The
outcome of simulating each trajectory is again a point $(t_{\varkappa },\ln
L_{\varkappa }^{i},Z_{\varkappa }).$

In the experiments we evaluate the expectation 
\begin{eqnarray}
\tilde{V}_{caplet}(0) &=&E^{\mathrm{Q}^{T_{i+1}}}\left[ \left(
L^{i}(T_{i})-K\right) _{+}\chi \left( \theta >T_{i}\right) +Z(\tau )\right]
\label{capl_appr} \\
&\approx &E^{\mathrm{Q}^{T_{i+1}}}\left[ \left( \exp (\ln L_{\varkappa
}^{i})-K\right) _{+}\chi \left( \varkappa =M\right) +Z_{\varkappa }\right] .
\notag
\end{eqnarray}%
The approximate equality in (\ref{capl_appr}) is related to the bias due to
the numerical approximation. The expectation on the right-hand side is
realized via the Monte Carlo technique.

\subsection{Numerical results}

Here we present some results of numerical tests of Algorithms~\ref{Hca01}
and~\ref{Hca02} for pricing the barrier caplet (\ref{capl_appr}). We use the
following parameters in the experiments: $i=9,$ $K=1\%,$ $H=28\%,$ $\delta
=1,$ $L^{9}(0)=13\%.$ The volatility $\sigma _{i}(t)$ is assumed to be
constant at $25\%$ . The exact caplet price with these parameters evaluated
by (\ref{caplet_exact}) is $6.57\%$. In the experiments we did $10^{6}$
Monte Carlo runs. The results are presented in Figure~\ref{Fig1}. We see
that Algorithm~\ref{Hca01} is much more accurate than Algorithm~\ref{Hca02}.
We also observe \textquotedblleft oscillating\textquotedblright\ convergence
which is typical for binary tree methods \cite{Der95}.

\FRAME{ftbhFU}{4.7591in}{3.5812in}{0pt}{\Qcb{\textit{Barrier caplet price: }%
Comparision of\textit{\ }the\textit{\ }results of numerical experiments for
the Algorithm~\protect\ref{Hca01} (Algorithm~$O(h)$) and~Algorithm~\protect
\ref{Hca02} (Algorithm $O(\protect\sqrt{h})$) and the exact caplet price
(solid line) evaluated for $i=9,$ $K=1\%,$ $H=28\%,$ $\protect\delta =1,$ $%
L^{9}(0)=13\%,$ $\protect\sigma _{i}(t)=25\%.$}}{\Qlb{Fig1}}{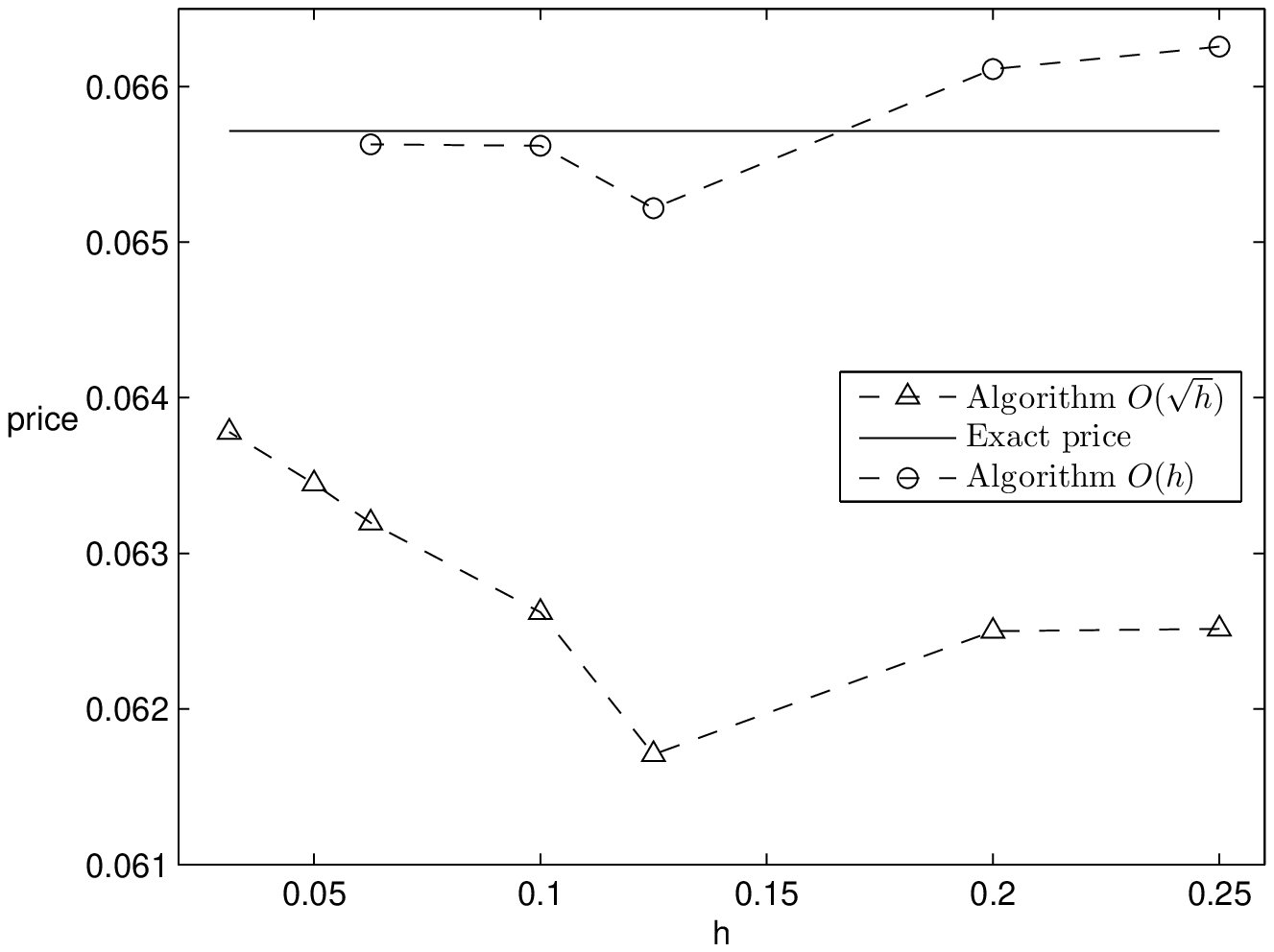}{%
\special{language "Scientific Word";type "GRAPHIC";maintain-aspect-ratio
TRUE;display "USEDEF";valid_file "F";width 4.7591in;height 3.5812in;depth
0pt;original-width 5.834in;original-height 4.3863in;cropleft "0";croptop
"1";cropright "1";cropbottom "0";filename 'caplet2.eps';file-properties
"XNPEU";}}

Let us also remark on the effect of variance reduction in these experiments.
For instance, in Algorithm~\ref{Hca01} for $h=0.02$ we got the Monte Carlo
error (i.e., half of the size of the confidence interval for corresponding
estimator with probability $0.95$) equal to $1.11\times 10^{-4}$ in the case
of $F=0$ and $1.55\times 10^{-5}$ in the case of the optimal $F$ from (\ref%
{fcap}) (i.e., $100$ time speed-up in reaching the same level of the Monte
Carlo error). The use of the optimal $F$ does not result in zero Monte Carlo
error due to the error of numerical integration.

\section{Trigger swap\label{sec_trig}}

This example is devoted to evaluation of multi-asset barrier options with
barriers on all or some of the correlated underlying assets. We consider a
trigger swap as a specific case, though the considered approach can be used
to value other multi-asset barrier options, for instance basket options,
CDOs and n$^{th}$-to-default CDSs, and it can also be applied for options
with single, double and time-dependent barriers.

A trigger swap is a swap on a floating reference rate that takes effect or
terminates when some index rate hits a specified trigger level. Trigger
swaps have a number of variations \cite{BrigoMercurio,Schoe}. Here we
consider a knock-in version of a payer trigger swap with a fixed rate $K$
whose barrier is continuously monitored. The index and reference rate both
coincide with a LIBOR rate. For given trigger levels $H^{0},\ldots ,H^{N-1}$
associated with the LIBOR rates $L^{0}(t),\ldots ,L^{N-1}(t),$ the structure
of the swap under consideration is expressed as follows. Once one of the the
continuously monitored LIBOR rate $L^{0}(t),\ldots ,L^{N-1}(t)$ for the
first time hits the corresponding trigger level $H^{0},\ldots ,H^{N-1}$ from
below, the contract holder enters into the payer swap starting at next tenor
date for the remaining time to the last tenor $T_{N-1}$. More specifically,
let $\theta $ be the first exit time of $L^{0}(s),\ldots ,L^{N-1}(s),$ $%
s\geq 0,$ from the domain $G=\left( 0,H^{0}\right) \times \cdots \times
\left( 0,H^{N-1}\right) ,$ $\tau $ be the first exit time of the space-time
diffusion $(s,L^{0}(s),\ldots ,L^{N-1}(s))$ from the domain $%
Q=[0,T_{N-1})\times G$ (clearly $\tau =\theta \wedge T_{N-1})$, and $%
T_{\varrho (\tau )}$ be the closest tenor date $T_{i}$ to $\tau $ from the
right, i.e., $\varrho (t)$ is defined as 
\begin{equation*}
\varrho (t)=\min \left\{ i,\text{ }i=0,1,\ldots ,N-1:t\leq T_{i}\right\} .
\end{equation*}%
If $\theta \leq T_{N-1},$ then at a tenor date $T_{\varrho (\tau )}$ the
contract holder enters into the contract according to which the holder pays
fixed payments of $\delta K$ and receives floating payments of $\delta
L^{i-1}(T_{i-1})$ at the coupon dates $T_{i},$ $i=\varrho (\tau )+1,\ldots
,N;$ otherwise the contract expires worthless.

The value of this trigger swap at time $t=0$ under the forward measure $%
\mathrm{Q}^{T_{N}}$ is given by%
\begin{eqnarray}
V_{trswap}(0) &=&P(0,T_{N})E^{\mathrm{Q}^{T_{N}}}\left[ \frac{1}{%
P(T_{\varrho (\tau )},T_{N})}\right.  \label{trswap} \\
&&\times \left. \left( 1-P(T_{\varrho (\tau )},T_{N})-K\delta
\dsum\limits_{i=\varrho (\tau )+1}^{N}P(T_{\varrho (\tau )},T_{i})\right)
\chi (\theta \leq T_{N-1})\right] ,  \notag
\end{eqnarray}%
or in terms of the LIBOR rates%
\begin{eqnarray}
V_{trswap}(0) &=&P(0,T_{N})E^{\mathrm{Q}^{T_{N}}}\left[ \left(
\prod\limits_{j=\varrho (\tau )}^{N-1}\left( 1+\delta L^{j}(T_{\varrho (\tau
)})\right) \right. \right.  \label{trswap_libor} \\
&&\left. \left. -K\delta \dsum\limits_{i=\varrho (\tau
)+1}^{N}\prod\limits_{j=i}^{N-1}\left( 1+\delta L^{j}(T_{\varrho (\tau
)})\right) -1\right) \chi (\theta \leq T_{N-1})\right] .  \notag
\end{eqnarray}

In order to price this contract by the Monte Carlo technique, we need to
generate paths for the vector $L^{\varrho (t)}(t),\ldots ,L^{N-1}(t)$. This
means that the size of the vector of the LIBOR rates which we need to
simulate decreases over time. The dynamics of $L^{i}(t)$ under $\mathrm{Q}%
^{T_{N}}$ are described by the SDEs (cf. (\ref{sde})):%
\begin{equation}
\frac{dL^{i}(t)}{L^{i}(t)}=-\sigma _{i}(t)\sum\limits_{j=i+1}^{N-1}\frac{%
\delta L^{j}(t)}{1+\delta L^{j}(t)}\rho _{i,j}\sigma _{j}(t)dt+\sigma
_{i}(t)dW_{i}^{T_{N}}(t),\text{ }i=\varrho (t),\ldots ,N-1.
\label{trswap_sde}
\end{equation}

\subsection{Algorithm\label{alg_trigger}}

Here we only apply Algorithm~\ref{Hca01}. For simplicity, we consider such
set of tenor dates $T_{i}$ and time steps $h$ that $T_{i}/h$ are integers.

As before, we simulate dynamics of the LIBOR rates $L^{\varrho
(t)}(t),\ldots ,L^{N-1}(t)$ in log space according to the weak Euler scheme
(cf. (\ref{Hc01})):%
\begin{eqnarray}
\ln L_{k+1}^{i} &=&\ln L_{k}^{i}-\sigma _{i}(t_{k})h\sum\limits_{j=i+1}^{N-1}%
\frac{\delta L_{k}^{j}}{1+\delta L_{k}^{j}}\rho ^{ij}\sigma _{j}(t_{k})
\label{treuler} \\
&&-\frac{1}{2}\left( \sigma _{i}(t_{k})\right) ^{2}h+\sigma _{i}(t_{k})\sqrt{%
h}\tsum\limits_{j=\varrho _{k}}^{N}U_{i,j}\xi _{j,k+1},  \notag \\
\text{ }i &=&\varrho _{k+1},\ldots ,N-1,\text{ }  \notag
\end{eqnarray}%
where $\xi _{j,k}$ are mutually independent random variables distributed by
the law $P(\xi =\pm 1)=1/2$ and $\varrho _{k}=:\varrho (t_{k}).$

The algorithm for the considered trigger swap proceeds as follows. Let $%
L_{k}=(L_{k}^{\varrho _{k}},\ldots ,L_{k}^{N-1})^{\top }.$ Denote by $%
\varkappa $ the first exit time of $(t_{k},L_{k})$ from $Q.$ Let $%
M=T_{N-1}/h.$ Suppose by time step $k$ none of the rates $L_{k}^{\varrho
_{k}},\ldots ,L_{k}^{N-1}$ have crossed their barriers $H^{\alpha
_{k}},\ldots ,H^{N-1},$ i.e., $\chi \left( \varkappa \leq k\right) =0$. Then
we evaluate whether at the next time step $k+1$ the event $\varkappa =k+1$
might be realized. One can see that the rate $L_{k+1}^{i},$ $i=\varrho
_{k+1},\ldots ,N-1,$ computed via (\ref{treuler}) will be below the barrier $%
H^{i},$ i.e. inside the domain $G$, if the following is true 
\begin{equation}
\ln L_{k}^{i}<\ln H^{i}-\lambda _{k}\sqrt{h}\ ,  \label{cond}
\end{equation}%
where 
\begin{equation*}
\lambda _{k}=\sigma _{Max}\sqrt{N-\varrho _{k}}
\end{equation*}%
and $\sigma _{Max}=\max_{,j,k}\sigma _{j}(t_{k}).$

If (\ref{cond}) is satisfied at time $t_{k}$ for all rates $L_{k}^{\varrho
_{k+1}},\ldots ,L_{k}^{N-1}$ then we move to step $t_{k+1},$ evaluate $\ln
L_{k+1}^{i},$ $i=\varrho _{k+1},\ldots ,N-1,$ according to (\ref{treuler})
and continue with the algorithm unless $k+1=M$ (in this case the trigger
swap expires worthless).

The case when the condition (\ref{cond}) does not hold for a single $i$
implies that the point $L_{k}$ is in the boundary zone $S_{t_{k},h}$ and is
near the barrier $H^{i}.$ Then we either assign $\varkappa =k,$ $\ln
L_{\varkappa }^{i}=\ln H^{i},\ \ln L_{\varkappa }^{j}=\ln L_{k}^{j}$ for $%
j\neq i,$ $T_{\varrho _{\varkappa }}=T_{\varrho _{k}}$ with probability 
\begin{equation}
p^{i}=\frac{\lambda _{k}\sqrt{h}}{\ln H^{i}-\ln L_{k}^{i}+\lambda _{k}\sqrt{h%
}}  \label{p_i}
\end{equation}%
and carry on with simulating $\ln L_{k+1}^{i},$ $i=\varrho _{k+1},\ldots
,N-1,$ according to (\ref{treuler}) starting from $\ln L_{\varkappa }$ until
time $T_{\varrho _{\varkappa }}=\min \left\{ T_{i}:t_{\varkappa }\leq T_{i}%
\text{, }i=0,1,\ldots ,N-1\right\} $ (the barriers are \textquotedblleft
removed\textquotedblright\ in simulating the remaining part of this
trajectory); or we jump outside the boundary zone $S_{t_{k},h}$ by changing
the $i^{th}$ component of $\ln L_{k}$ from $\ln L_{k}^{i}$ to $\ln
L_{k}^{i}-\lambda _{k}\sqrt{h}$ with probability $1-p^{i},$ perform the
usual step according to (\ref{treuler}) and continue with the algorithm
unless $k+1=M$ (in this case the trigger swap expires worthless).

We note that in comparison with the original formulation of Algorithm~\ref%
{Hca01} here we do not stop the chain $L_{k}$ at its first exit time from
the space domain $G.$ Instead, when the barrier is hit, we find the trigger
tenor date $T_{\varrho _{\varkappa }},$ and if $t_{\varkappa }<T_{N-1},$ we
continue the simulation according to (\ref{treuler}) until $T_{\varrho
_{\varkappa }}$ to get the required LIBOR rates $L^{j}(T_{\varrho (\tau )})$
in (\ref{trswap_libor}).

Now, we discuss the case when the condition (\ref{cond}) does not hold for
more than one $i$ (i.e., the random walk has reached a corner of the domain $%
G).$ In this case, the algorithm proceeds as follows. Let us denote by $\ell
=\left\{ l_{1},\ldots ,l_{n}\right\} $ the set of tenor dates corresponding
to the LIBOR rates for which (\ref{cond}) is violated. First, we select the
rate from the set $\left\{ \ln L_{k}^{l_{1}},\ldots ,\ln
L_{k}^{l_{n}}\right\} ,$ which is the closest to its boundary, i.e. $l_{j}$
such that $\ln H^{l_{j}}-\ln L_{k}^{l_{j}}$ is minimum over $j=1,\ldots ,n.$
Then, we repeat the procedure which is given above for the single $i$ with
the following difference. If $\ln L_{k}^{i}$ jumps from the boundary to $\ln
L_{k}^{i}-\lambda _{k}\sqrt{h},$ we find the second closest rate from the
set $\left\{ \ln L_{k}^{l_{1}},\ldots ,\ln L_{k}^{l_{n}}\right\} $ and as
before repeat for this point the routine we have presented for the single $i.
$ We follow this procedure in the outlined fashion until either the set $%
\ell $ is empty or for some $l_{j\text{ }}$ we reach the boundary and assign 
$\ln L_{k}^{l_{j\text{ }}}=\ln H^{l_{j\text{ }}}.$

The outcome of simulating each trajectory is the payer swap starting tenor
date $T_{\varrho _{\varkappa }},$ the stopping time $\varkappa $ and the
point $\ln L_{\eta }$ with $\eta =T_{\varrho _{\varkappa }}/h,$ which are
used for evaluating the trigger swap: 
\begin{eqnarray*}
V_{trswap}(0) &\approx &P(0,T_{N})E^{\mathrm{Q}^{T_{N}}}\left[ \left(
\prod\limits_{j=\varrho _{\varkappa }}^{N-1}\left( 1+\delta L_{\eta
}^{j}\right) \right. \right. \\
&&\left. \left. -K\delta \dsum\limits_{i=\varrho _{\varkappa
}+1}^{N}\prod\limits_{j=i}^{N-1}\left( 1+\delta L_{\eta }^{j}\right)
-1\right) \chi (\varkappa <M)\right]
\end{eqnarray*}%
with the expectation simulated by the Monte Carlo technique. We present the
pseudocode for simulating a single trajectory based on the algorithm we
described above.

{%
\begin{algorithm}[htb]
\caption{Pseudocode for simulating a single trajectory in pricing the barrier trigger swap}
\label{alg:algorithm1}
{\small
SET  $M$  to $T_{N-1}/h,$ $k$ to $1$, $\kappa$ to $M$

WHILE $k<M$

\qquad IF $\kappa>k$

\qquad \qquad FOR $j=0$ to $N-1,$ 

\qquad \qquad \qquad IF (\ref{cond}) for $j$
is false

\qquad \qquad \qquad \qquad calculate $p^{j}$ by  (\ref{p_i})

\qquad \qquad \qquad \qquad form array $p$ of $p^{j}$

\qquad  \qquad \qquad ENDIF

\qquad \qquad ENDFOR

 \qquad \qquad  sort $p$ in descending order

 \qquad \qquad FOR $n=1$ to $length(p)$ 

\qquad \qquad \qquad generate $u\sim Unif[0.1];$

\qquad  \qquad \qquad \qquad IF $u<p(n)$

\qquad\qquad \qquad \qquad \qquad SET $\kappa$ to $k$

\qquad \qquad \qquad \qquad \qquad SET $\ln L_{k}^{i}$ to $\ln H^{i}$

\qquad \qquad \qquad \qquad \qquad SET $M$ to $T_{\varrho_{\kappa} }/h$

\qquad \qquad \qquad \qquad \qquad BREAK

\qquad \qquad \qquad \qquad ELSE

\qquad \qquad \qquad \qquad \qquad SET $\ln L_{k}^{i}$ to $\ln L_{k}^{i}-\lambda_k%
\sqrt{h}$

\qquad \qquad \qquad \qquad ENDIF

\qquad \qquad ENDFOR

\qquad ENDIF

\qquad Evaluate $\ln L_{k+1}$ by (\ref{treuler}%
)

\qquad Increase $k$ by $1$

ENDWHILE
}
\end{algorithm}}

\subsection{Numerical results}

Let us present results of numerical experiments we performed for pricing a
trigger swap with Algorithm$~$\ref{alg_swaption}. The parameters chosen for
the experiments are $T^{0}=5,$ $T^{\ast }=16,\ K=0.01,$ $H=0.13,$ $\delta
=1, $ $\beta =0.2.$ The initial LIBOR rate curve is assumed to be flat at $%
4\%$ and the volatility $\sigma _{i}(t)$ is set to be constant at $20\%$ .
In the simulations we run $10^{6}$ Monte Carlo paths.

Since the closed-form formula for trigger swap (\ref{trswap_libor}) is not
available, we found the reference trigger swap price by evaluating the price
using Algorithm~\ref{alg_trigger} with $h=0.01$ and the number of Monte
Carlo runs $10^{6}$. This reference price is $5.46\times 10^{-2}$ with the
Monte Carlo error $5.50\times 10^{-4},$ which gives half of the size of the
confidence interval for the corresponding estimator with probability $0.95$.

The results of the experiments with Algorithm~\ref{alg_trigger} are
presented in Table~\ref{tab2}. In the table, the values before
\textquotedblleft $\pm $\textquotedblright\ are estimates of the bias
computed as the difference between the reference price and its sampled
approximation, while the values after \textquotedblleft $\pm $%
\textquotedblright\ give half of the size of the confidence interval for the
corresponding estimator with probability $0.95$. The \textquotedblleft mean
exit time\textquotedblright\ is the average time for trajectories $%
(t_{k},L_{k})$ to leave the space-time domain $Q$. The experimentally
observed convergence rate for Algorithm~\ref{alg_trigger} is in agreement
with the theoretical first order convergence in $h$ (though we note that the
convergence theorem in \cite{MT02,MT1} is proved under restrictive
regularity conditions and the payoff of the trigger swap and the boundary of
the space domain $G$ do not satisfy these conditions).

\begin{table}[h] \centering%
\caption{{\it Performance of Algorithm~5.1 for the trigger swap.}
\label{tab2}}
\setlength{\tabcolsep}{2pt} 
\begin{tabular}{ccc}
\hline
$h$ & error & $\ \ \ \ $mean exit time$\ \ \ \ $ \\ \hline
\multicolumn{1}{l}{$\ \ \ \ 0.25\ \ \ \ $} & $2.22\times 10^{-2}\pm
6.39\times 10^{-4}$ & $\ \ \ \ \ \ 12.51\ \ \ \ \ \ $ \\ \hline
\multicolumn{1}{l}{$\ \ \ \ 0.2\ \ \ \ $} & $1.85\times 10^{-2}\pm
6.26\times 10^{-4}$ & $12.61$ \\ \hline
\multicolumn{1}{l}{$\ \ \ \ 0.125\ \ \ \ $} & $1.17\times 10^{-2}\pm
6.01\times 10^{-4}$ & $12.78$ \\ \hline
\multicolumn{1}{l}{$\ \ \ \ 0.1\ \ \ \ $} & $9.56\times 10^{-3}\pm
5.92\times 10^{-4}$ & $12.83$ \\ \hline
\multicolumn{1}{l}{$\ \ \ \ 0.0625\ \ \ \ $} & $6.03\times 10^{-3}\pm
5.78\times 10^{-4}$ & $12.92$ \\ \hline
$0.05\ \ \ \ $ & $4.67\times 10^{-3}\pm 5.72\times 10^{-4}$ & $12.95$ \\ 
\hline
\end{tabular}%
\end{table}%

\section{Barrier swaption\label{sec_swpt}}

In this section we consider Monte Carlo evaluation of a knock-out swaption
under the LMM. We use the knock-out swaption as a guide in our exposition,
its treatment is rather general and it can be used to value different
barrier options, where the underlying and barrier can be expressed as
functionals of some diffusion process.

A European payer (receiver) swaption is an option that gives its holder a
right, but not an obligation, to enter a payer (receiver) swap at a future
date at a given fixed rate $K$. Usually, the swaption maturity coincides
with the first reset date $T_{0}$ of the underlying swap. The underlying
swap length $T_{N}-T_{0}$ is called the tenor of the swaption.

Without loss of generality, we concentrate on a knock-out receiver swaption
with the first reset date $T_{0}$. A knock-out swaption has the structure as
a standard swaption except that if the underlying swap rate is above a
barrier level $R_{up}$ at any time before $T_{0}$ then the swaption expires
worthless. The price of the knock-out swaption at time $t=0$ under the
forward measure $\mathrm{Q}^{T_{0}}$ is given by:%
\begin{equation}
V_{swaption}(0)=P(0,T_{0})E^{\mathrm{Q}^{T_{0}}}\left[ \delta \left(
R_{swap}(T_{0})-K\right) _{+}\dsum\limits_{j=1}^{N}P(T_{0},T_{j})\chi \left(
\theta >T_{0}\right) \right] ,  \label{swaption price}
\end{equation}%
where $\theta $ is the first exit time of the process $R_{swap}(s),$ $s\geq
0,$ from the interval $(0,R_{up}).$ The swap rate $R_{swap}(s)$ can be
expressed in terms of the spanning LIBOR rates as%
\begin{equation}
R_{swap}(s)=\frac{1-1/\prod\limits_{j=0}^{N-1}\left( 1+\delta
L^{j}(s)\right) }{\delta
\dsum\limits_{i=0}^{N-1}1/\prod\limits_{j=0}^{i}\left( 1+\delta
L^{j}(s)\right) }\ .  \label{swap_rate}
\end{equation}%
The bond prices $P(T_{0},T_{j})$ can also be expressed via LIBOR rates (see (%
\ref{libor_bond})).

Also, let $\tau $ be the first exit time of the space-time process $%
(s,R_{swap}(s))$ from the domain $D=[0,T_{0})\times (0,R_{up})$ (obviously, $%
\tau =\theta \wedge T_{0})$.

We note that expression (\ref{swaption price}) depends on the joint
distribution of the forward rates $L^{0}(T_{0}),\ldots ,L^{N-1}(T_{0}).$ The
LMM dynamics of LIBOR rates under $\mathrm{Q}^{T_{0}}$ are given by (cf. (%
\ref{sde})): 
\begin{equation}
\frac{dL^{i}(t)}{L^{i}(t)}=\sigma _{i}(t)\sum\limits_{j=0}^{i}\frac{\delta
L^{j}(t)}{1+\delta L^{j}(t)}\rho _{i,j}\sigma _{j}(t)dt+\sigma
_{i}(t)dW_{i}^{T_{0}}(t),\text{ }i=0,\ldots ,N-1.  \label{sde swaption}
\end{equation}

In this example we deal with pricing the barrier swaption (\ref{swaption
price}) expressed in terms of the spanning LIBOR rates with dynamics in the
form of (\ref{sde swaption}). This means that we consider this problem in
the coordinate system of the spanning LIBOR rates and the barrier is given
as an implicit surface in the LIBOR coordinates. We also note that there is
the space domain $G$ in the phase space of the SDEs (\ref{sde swaption})
corresponding to the interval $(0,R_{up})$ on the swap-rate semi-line. As
usual, the corresponding space-time domain $Q:=[0,T_{0})\times G.$

For test purposes, let us introduce an analytical approximation for the
barrier swaption. To this end, we note that under the Swap Market Model
(SMM, see details in \cite{BrigoMercurio,Rebo,Schoe}) the barrier swaption
pricing problem admits the closed-form solution (cf. (\ref{caplet_exact}))%
\begin{eqnarray}
&&V_{swaption}(0)=\delta \dsum\limits_{j=1}^{N}P(0,T_{j})\left\{ R_{swap}(0) 
\left[ \Phi (\delta _{+}(R_{swap}(0)/K,v_{R_{swap}}))\right. \right.
\label{swaption exact} \\
&&\left. -\Phi (\delta _{+}(R_{swap}(0)/R_{up},v_{R_{swap}}))\right]  \notag
\\
&&-K\left[ \Phi (\delta _{-}(R_{swap}(0)/K,v_{R_{swap}}))-\Phi (\delta
_{-}(R_{swap}(0)/R_{up},v_{R_{swap}}))\right]  \notag \\
&&-H\left[ \Phi (\delta _{+}(R_{up}^{2}/(KR_{swap}(0)),v_{R_{swap}}))-\Phi
(\delta _{+}(R_{up}/R_{swap}(0),v_{R_{swap}}))\right]  \notag \\
&&+KR_{swap}(0)\Phi (\delta
_{-}(R_{up}{}^{2}/(KR_{swap}(0)),v_{R_{swap}})/R_{up})  \notag \\
&&\left. \left. -\Phi (\delta _{-}(R_{up}/R_{swap}(0),v_{R_{swap}}))\right]
\right\} ,  \notag
\end{eqnarray}%
where $\delta _{\pm }$ are from (\ref{d1}),%
\begin{equation*}
v_{R_{swap}}^{2}=\int_{0}^{T_{i}}\left( \sigma _{R_{swap}}(s)\right) ^{2}ds,
\end{equation*}%
and $\sigma _{R_{swap}}(s)$ is the instantaneous volatility of the
log-normal dynamics of the swap rate.

Using Rebonato's formula \cite[p. 283]{BrigoMercurio}, we can approximately
compute the \textquotedblleft approximate\textquotedblright\ volatility $%
v_{R_{swap}}^{LMM}$ for the LMM analogous to the volatility $v_{R_{swap}}$
in the SMM entering (\ref{swaption exact}) as%
\begin{equation}
v_{R_{swap}}^{LMM}=\sum\limits_{i,j=0}^{N-1}\frac{\omega _{i}(0)\omega
_{j}(0)L^{i}(0)L^{j}(0)\rho _{ij}}{\left( R_{swap}(0)\right) ^{2}}%
\int_{0}^{T_{0}}\sigma _{i}(s)\sigma _{j}(s)ds,  \label{rebo}
\end{equation}%
where%
\begin{equation*}
\omega _{i}(0)=\frac{1-1/\prod\limits_{j=0}^{i-1}\left( 1+\delta
L^{j}(0)\right) }{\delta
\dsum\limits_{k=0}^{N-1}1/\prod\limits_{j=0}^{k}\left( 1+\delta
L^{j}(0)\right) }.
\end{equation*}

The quantity $v_{R_{swap}}^{LMM}$ can be used as a proxy for $v_{R_{swap}}$
in (\ref{swaption exact}) to compute approximated barrier swaption prices
under LMM. We will check in our numerical experiments whether an
approximation obtained by our algorithm is consistent with this analytical
approximation.

\subsection{Algorithm\label{alg_swaption}}

Here we exploit Algorithm~\ref{Hca01}. We choose a time step $h>0$ so that $%
M=T_{0}/h$ is an integer. Again inside the domain $G$ we use the weak Euler
scheme to simulate trajectories of the log LIBOR rates (\ref{sde swaption}):%
\begin{eqnarray}
\ln L_{k+1}^{i} &=&\ln L_{k}^{i}+\sigma _{i}(t_{k})\sum\limits_{j=0}^{i}%
\frac{\delta L_{k}^{j}}{1+\delta L_{k}^{j}}\rho _{ij}\sigma _{j}(t_{k})h
\label{sweuler} \\
&&-\frac{1}{2}\left( \sigma _{i}(t_{k})\right) ^{2}h+\sigma _{i}(t_{k})\sqrt{%
h}\tsum\limits_{j=0}^{N-1}U_{i,j}\xi _{j,k+1},  \notag \\
\text{ }i &=&0,\ldots ,N-1,\text{ }  \notag
\end{eqnarray}%
where $\xi _{j,k}$ are mutually independent random variables distributed by
the law $P(\xi =\pm 1)=1/2.$

For a fixed $t_{k},$ we denote by $\ln L_{k}$ the point with coordinates $%
\ln L_{k}^{0},\ln L_{k}^{1},\ldots ,$ $\ln L_{k}^{N-1},$ i.e. $\ln
L_{k}=(\ln L_{k}^{0},\ln L_{k}^{1},\ldots ,\ln L_{k}^{N-1})^{\top }.$ As
before, we follow the random walk constructed by (\ref{sweuler}) until we
reach the boundary zone $S_{t_{k},h}.$ Algorithmically, it implies that we
implement a check at each step whether the current position of the random
walk is in the boundary zone $S_{t_{k},h}$. More precisely, we evaluate at
time $t_{k}$ whether the current position $\ln L_{k}$ is such that the
maximum increment from point $\ln L_{k}$ according to all possible
realizations of (\ref{sweuler}) at the next time level $t_{k+1}$ results in
the state of the random walk below the barrier, i.e. in the domain $G.$

Introduce%
\begin{equation*}
\ln L_{k,Max}=\max_{i}\ln L_{k}^{i}\ 
\end{equation*}%
and%
\begin{equation}
\ln \hat{L}_{k+1}=\ln L_{k,Max}+\sigma _{Max}^{2}hN-\frac{1}{2}\sigma
_{Max}^{2}h+\sigma _{Max}\sqrt{hN},  \label{L hat}
\end{equation}%
where $\sigma _{Max}=\max_{i,k}\sigma _{i}(t_{k})$. Using the fact that%
\begin{equation*}
R_{swap}(\hat{L}_{k+1},...,\hat{L}_{k+1})=\hat{L}_{k+1},
\end{equation*}%
one can see that the current position of the random walk $\ln L_{k}$ is
inside the domain $G$ if the following condition is satisfied 
\begin{equation}
\ln \hat{L}_{k+1}<\ln R_{up}.  \label{rough cond}
\end{equation}%
Algorithmically, we do the following. If condition (\ref{rough cond}) is
true, we evaluate the next position of the random walk at $t_{k+1}$
according to (\ref{sweuler}) and continue further with the algorithm unless $%
k+1=M$ (i.e., we have reached the maturity time $T_{0}$ of the swaption).

We note the condition (\ref{rough cond}) is computationally easy to evaluate
but it is rather rough. Once this condition fails, we check a finer but
computationally more expensive condition based on the maximum increments
from each of $L^{i}(t_{k})$ towards the boundary:%
\begin{gather}
R_{swap}(L_{k}^{0}(1+\sigma _{0}(t_{k})\sigma _{Max,k}h+\sigma _{0}(t_{k})%
\sqrt{Nh}),L_{k}^{1}(1+2\sigma _{1}(t_{k})\sigma _{Max,k}h  \notag \\
+\sigma _{1}(t_{k})\sqrt{(N-1)h}),...,L_{k}^{N-1}(1+N\sigma
_{N-1}(t_{k})\sigma _{Max,k}h+\sigma _{N-1}(t_{k})\sqrt{h}))<R_{up},
\label{fine cond}
\end{gather}%
where $\sigma _{Max,k}=\max_{j}\sigma _{j}(t_{k}).$ If condition (\ref{fine
cond}) holds, we again carry on to the next time step $t_{k+1}$ using (\ref%
{sweuler}) and continue further with the algorithm unless $k+1=M.$

If both conditions (\ref{rough cond}) and (\ref{fine cond}) fail, the random
walk has reached the boundary zone $S_{t_{k},h},$ where as before we apply
the different procedure which require us to find the projection $\ln
L_{k}^{\pi }:=(L_{k}^{\pi ,0},L_{k}^{\pi ,1},\ldots ,L_{k}^{\pi ,N-1})^{\top
}$ of the current position $\ln L_{k}$ on the boundary given as the implicit
function of the spanning LIBOR rates:%
\begin{equation}
\ln R_{swap}(t_{k})=\ln R_{up}.  \label{barrier}
\end{equation}

For completeness of the exposition, let us discuss how the projection $\ln
L_{k}^{\pi }$ can be simulated before we return to the description of the
algorithm. The problem of finding point $\ln L_{k}^{\pi }$ is equivalent to
finding the minimum value of the function 
\begin{equation}
|\ln L_{k}^{\pi }-\ln L_{k}|^{2}=\left( \ln L_{k}^{\pi ,0}-\ln
L_{k}^{0}\right) ^{2}+\cdots +\left( L_{k}^{\pi ,N-1}-\ln L_{k}^{N-1}\right)
^{2}  \label{QP}
\end{equation}%
subject to the constraint%
\begin{equation}
\ln \left( \frac{\prod\limits_{j=0}^{N-1}\left( 1+\delta L_{k}^{\pi
,j}\right) -1}{\delta \left(
1+\dsum\limits_{i=0}^{N-2}\prod\limits_{j=i+1}^{N-1}\left( 1+\delta
L_{k}^{\pi ,j}\right) \right) }\right) =\ln R_{up}.  \label{constrain}
\end{equation}%
We regard $\ln L_{k}^{\pi ,1},\ldots ,\ln L^{\pi ,N-1}$ as independent
variable in the constraint equation (\ref{constrain}) and write $\ln
L_{k}^{\pi ,0}$ as 
\begin{equation}
\ln L_{k}^{\pi ,0}=\ln \left( \frac{R_{up}\cdot \left(
1+\dsum\limits_{i=0}^{N-2}\prod\limits_{j=i+1}^{N-1}\left( 1+\delta
L_{k}^{\pi ,j}\right) \right) +1}{\prod\limits_{j=1}^{N-1}\left( 1+\delta
L_{k}^{\pi ,j}\right) }-\frac{1}{\delta }\right) .  \label{L_0}
\end{equation}%
Hence the minimization problem is reduced to finding the point $\ln
L_{k}^{\pi ,1},$ $\ldots ,\ln L^{\pi ,N-1}$ at which the function $|\ln
L_{k}^{\pi }-\ln L_{k}|^{2}$ from (\ref{QP}) with $\ln L_{k}^{\pi ,0}$ from (%
\ref{L_0}) has its minimum value. This optimization problem can be solved
using standard procedures, e.g. the MATLAB function \textquotedblleft 
\textrm{lsqnonlin()}\textquotedblright .

Let us now continue with the description of the algorithm. When $\ln L_{k}$
is in the boundary zone, we either stop the chain at $\ln L_{k}^{\pi }$ with
probability $p:$ 
\begin{equation}
p=\frac{\lambda \sqrt{h}}{|\ln L_{k}^{\pi }-\ln L_{k}|+\lambda \sqrt{h}},
\label{p_sw}
\end{equation}%
where 
\begin{equation}
\lambda \sqrt{h}=\sqrt{N}\left( \sigma _{Max}^{2}hN-\frac{1}{2}\sigma
_{Max}^{2}h+\sigma _{Max}\sqrt{hN}\right) ;  \label{lamda_sw}
\end{equation}%
or we jump inside the domain $Q$ to the point $\ln L_{k}+\lambda \sqrt{h}%
\frac{\overrightarrow{\ln L_{k}^{\pi }\ \ln L_{k}}}{|\ln L_{k}^{\pi }-\ln
L_{k}|}$ with probability $1-p,$ apply the Euler step (\ref{sweuler}) to
evaluate $\ln L_{k+1}$ and continue further with the algorithm unless $%
k+1=M. $

The outcome of simulating each trajectory is the point $(t_{\varkappa },\ln
L_{\varkappa })$. In Algorithm~\ref{alg2}\ we present the pseudocode for
simulating a single trajectory based on the algorithm we described above.

\begin{algorithm}[thb]
\caption{Barrier swaption: Pseudocode for simulating a single trajectory}
\label{alg2}

{\small FOR }$k=1${\small \ to }$M$

{\small \qquad calculate }$\ln \hat{L}_{k+1}${\small \ by (\ref{L hat})}

{\small \qquad IF (\ref{rough cond}) is true: }
{\small  calculate }$\ln L_{k+1}${\small \ by (\ref%
{sweuler})}

{\small \qquad ELSE }

{\small \qquad \qquad IF (\ref{fine cond}) is true:}
{\small calculate }$\ln L_{k+1}${\small \ by (\ref%
{sweuler})}

{\small \qquad \qquad ELSE }

{\small \qquad \qquad \qquad solve the minimisation problem (\ref{QP})
with $\ln L_{k}^{\pi,0}$ from (\ref{L_0})}

{\small \qquad \qquad \qquad generate }$u\sim Unif[0,1]$

{\small \qquad \qquad  \qquad calculate probability }$p${\small \ by (\ref{p_sw})}

{\small \qquad \qquad \qquad IF }$u<p$

{\small \qquad \qquad \qquad \qquad break}

{\small \qquad \qquad \qquad  ELSE}

{\small \qquad \qquad \qquad \qquad  SET }$\ln L_{k}${\small \ to }$\ln L_{k}+\lambda \sqrt{h} \frac{%
\overrightarrow{\ln L_{k}^{\pi} \,  \ln L_{k}}}{\left\vert \ln L_{k}^{\pi}- \ln L_{k} \right\vert }$

{\small \qquad \qquad \qquad \qquad calculate }$\ln L_{k+1}${\small \ by
(\ref{sweuler})}

{\small \qquad \qquad \qquad ENDIF }

{\small \qquad \qquad ENDIF}

{\small \qquad ENDIF}

{\small ENDFOR}

\end{algorithm}

\subsection{Numerical results}

We give some results for pricing a barrier swaption by Algorithm$~$\ref%
{alg_swaption}. We consider the barrier swaption with the initial LIBOR
curve flat at $5\%,$ constant volatility $\sigma _{i}(t)$ at $10\%$ and the
following parameters: $T_{0}=10,$ $T^{\ast }=20,\ K=0.01,$ $R_{up}=0.075,$ $%
\delta =1,$ $\beta =0.1.$ The simulations use $10^{6}$ Monte Carlo runs.

The pricing problem for the barrier swaption (\ref{swaption price}) does not
admit a closed-form solution. We used the barrier swaption price found with
Algorithm~\ref{alg_swaption} with $h=0.01$ and $10^{7}$ of Monte Carlo runs
as the reference solution. This reference price is $0.15506$ with the Monte
Carlo error $1.42\times 10^{-4},$ which gives half of the size of the
confidence interval for the corresponding estimator with probability $0.95$.
The analytical approximation based on (\ref{swaption exact}) and (\ref{rebo}%
) yields the price of the barrier swaption $0.15556.$

We present results of the experiments in Table \ref{tab3}. As in the
previous section, the error column values before \textquotedblleft $\pm $%
\textquotedblright\ are estimates of the bias computed using the reference
price value and the values after \textquotedblleft $\pm $\textquotedblright\
reflect the Monte Carlo error with probability $0.95$. The \textquotedblleft
mean exit time\textquotedblright\ is the average time for approximate
trajectories to exit the space-time domain $Q$. It is clear that the results
demonstrate the expected first order of convergence.

\begin{table}[h] \centering%
\caption{{\it Performance of Algorithm~6.1 for the barrier swaption.}
\label{tab3}}
\setlength{\tabcolsep}{2pt} 
\begin{tabular}{ccc}
\hline
$h$ & error & $\ \ \ $mean$\ $exit time $\ \ \ \ $ \\ \hline
\multicolumn{1}{l}{$\ \ \ \ 0.25\ \ \ \ $} & $1.01\times 10^{-2}\pm
4.33\times 10^{-4}$ & $\ \ \ \ \ \ 9.36$ $\ \ \ \ \ \ $ \\ \hline
\multicolumn{1}{l}{$\ \ \ \ 0.2\ \ \ \ $} & $8.08\times 10^{-3}\pm
4.37\times 10^{-4}$ & $9.40$ \\ \hline
\multicolumn{1}{l}{$\ \ \ \ 0.125\ \ \ \ $} & $5.15\times 10^{-3}\pm
4.42\times 10^{-4}$ & $9.46$ \\ \hline
\multicolumn{1}{l}{$\ \ \ \ 0.1\ \ \ \ $} & $4.15\times 10^{-3}\pm
4.44\times 10^{-4}$ & $9.48$ \\ \hline
\multicolumn{1}{l}{$\ \ \ \ 0.0625\ \ \ \ $} & $2.58\times 10^{-3}\pm
4.47\times 10^{-4}$ & $9.51$ \\ \hline
$\ \ \ \ 0.03125\ \ \ \ $ & $1.03\times 10^{-3}\pm 4.49\times 10^{-4}$ & $%
9.54$ \\ \hline
\end{tabular}%
\end{table}%

\section*{Acknowledgment}

MVT was partially supported by the Leverhulme Trust.

\end{document}